\documentclass[fleqn]{article}
\usepackage{longtable}
\usepackage{graphicx}

\begin {document}
\begin{flushleft}
{\LARGE
{\bf Energy levels and radiative rates  for transitions in \\Ti VII}
}\\

\vspace{1.5 cm}

{\bf {Kanti  M  ~Aggarwal and Francis  P   ~Keenan}}\\ 

\vspace*{1.0cm}

Astrophysics Research Centre, School of Mathematics and Physics, Queen's University Belfast, Belfast BT7 1NN, Northern Ireland, UK\\ 
\vspace*{0.5 cm} 

e-mail: K.Aggarwal@qub.ac.uk \\

\vspace*{1.50cm}

Received  11 July 2013\\
Accepted for publication 1 November 2013 \\
Published xx November 2013 \\
Online at stacks.iop.org/PhysScr/vol/number \\

\vspace*{1.5cm}

PACS Ref: 31.25 Jf, 32.70 Cs,  95.30 Ky

\vspace*{1.0 cm}

\hrule

\vspace{0.5 cm}
{\Large {\bf S}} This article has associated online supplementary data files \\
Tables 4 and 5 are available only in the electronic version at stacks.iop.org/PhysScr/vol/number/mmedia

\end{flushleft}

\clearpage


\begin{abstract}

We report calculations of energy levels, radiative rates, oscillator strengths and line strengths for transitions among the lowest 231 levels of Ti VII.  The general-purpose relativistic atomic structure package ({\sc grasp}) and flexible atomic code ({\sc fac}) are adopted for the calculations. Radiative rates, oscillator strengths and line strengths are provided for all electric dipole (E1), magnetic dipole (M1), electric quadrupole (E2) and magnetic quadrupole (M2) transitions among the 231 levels, although calculations have been performed for a much larger number of levels (159,162). In addition, lifetimes for all 231 levels are listed. Comparisons are made with existing results and the accuracy of the data is assessed. In particular, the most recent calculations reported  by Singh {\em et al} [Can J. Phys. {\bf 90} (2012) 833] are found to be unreliable, with discrepancies for energy levels of up to 1 Ryd and for radiative rates of up to five orders of magnitude for several transitions, particularly the weaker ones. Based on several comparisons  among a variety of calculations with two independent codes, as well as with the earlier results, our listed energy levels are estimated to be accurate to better than 1\% (within 0.1 Ryd), whereas results for radiative rates and other related parameters  should be accurate to  better than 20\%. 
 
\end{abstract}

\clearpage

\section{Introduction}

Titanium, one of the iron group elements (Sc -- Zn), is an important constituent of fusion plasmas, and in particular it has  been used to reducing impurity concentrations and controlling hydrogen recycling \cite{bs}. Due to the high temperatures  in fusion plasmas, many   ionisation stages of titanium are observed. For modelling these plasmas, and particularly to assess the radiation losses, atomic data are required for several parameters, such as energy levels and  oscillator strengths or radiative decay rates. The need for atomic data has become even greater with the developing  ITER project. Therefore, in the recent past we have reported atomic data for several Ti ions, namely Ti VI \cite{ti6}, Ti X \cite{ti10}, Ti XIX \cite{ti19}, Ti XX \cite{ti20}, Ti XXI \cite{ti21} and Ti XXII  \cite{ti22}. In this paper  we focus our attention on S-like Ti VII.

Titanium may  also be important for the analysis of astrophysical plasmas,  with  several emission lines listed in the CHIANTI \cite{chianti} database at ${\tt {\verb+http://www.chiantidatabase.org+}}$. A coronal line of Ti VII ([3p$^4$] $^3$P$_2$--$^3$P$_1$) has been identified by Rudy {\em et al} \cite{rjr} in  Nova Cassiopeiae 1995, and they  also showed that the ratio of the $^3$P$_0$ -- $^1$D$_2$ and $^3$P$_2$ -- $^3$P$_1$ lines can be a useful density diagnostic at temperatures around 10,000 K. Similarly,  many emission lines are listed in the 125 -- 700 ${\rm \AA}$ wavelength range in the {\em Atomic Line List} (v2.04) of Peter van Hoof at ${\tt {\verb+http://www.pa.uky.edu/~peter/atomic/+}}$, because these are useful in the generation of synthetic spectra. Furthermore, laboratory measurements of Ti VII   lines have been made in laser-produced plasmas by Ebrahim {\em et al} \cite{nae}, and radiative lifetimes for two levels have been experimentally determined  by Dumont {\em et al} \cite{pdd} via the beam-foil spectroscopy method. More recently, Clementson {\em et al} \cite{jht} have measured intensities of emission lines in the 240 -- 370 ${\rm \AA}$ range from multiply charged Ti ions, including Ti VII, through the Sustained Spheromak Physics Experiment at the Lawrence Livermore National Laboratory. 

Available  experimental data for energy levels have been  compiled by  the NIST (National Institute of Standards and Technology) team \cite{sc}, \cite{yuri}. Considering the importance of Ti ions, several calculations have been performed for  Ti VII -- see for example Singh {\em et al} \cite{mm4} and references therein. However, two important ones  are by Bi{\`{e}}mont \cite{eb} and Froese-Fischer {\em et al} \cite{cff1}. They have  adopted the Hartree-Fock relativistic (HFR)  and the multi configuration Hartree-Fock (MCHF) methods, respectively, but have performed only limited calculations. Bi{\`{e}}mont calculated energies for 57 levels of the 3p$^4$, 3s3p$^5$, 3p$^3$3d and 3p$^3$4s configurations, whereas Froese-Fischer {\em et al} included  40 of the 47 levels of the first three configurations. Both calculations also report radiative rates (A- values) for some (not all) electric dipole (E1) transitions. However, for plasma modelling a complete set of atomic data is required, and preferably for a larger number of levels due to possible  cascading effect. In between  levels of the 3p$^3$3d and 3p$^3$4s configurations lie several of the 3s3p$^4$3d configuration (see current Table 4, available online at stacks.iop.org/PhysScr/vol/number/mmedia). These missing levels in the earlier calculations may affect any plasma modelling and particularly the determination of lifetimes.

To address the limitation in earlier work, Singh {\em et al} \cite{mm4}  have recently performed a larger calculation containing 114 levels, belonging to the 3p$^4$, 3s3p$^5$, 3p$^3$3d, 3p$^6$, 3s3p$^4$3d  and 3s3p4s configurations.  They have adopted the CIV3 code of Hibbert \cite{civ3} and have also included one-body relativistic operators in their calculations, which should be sufficient for a moderately heavy ion such as Ti VII. Furthermore, they included CI (configuration interaction) with up to 4$\ell$ orbitals.  In addition, they {\em adjusted} the Hamiltonian in accordance with the NIST compilations (a process known as ``fine-tuning"), which minimises the differences between theoretical and experimental energy levels.   However, their reported results remain incomplete for several reasons. First, to span all levels  of the 3p$^3$4s configuration one needs to include a minimum of 137 levels, because several levels of the 3p$^2$3d$^2$ and 3p$^3$4p configurations lie in between  (see current Table 4, available online at stacks.iop.org/PhysScr/vol/number/mmedia), which have not been calculated by  Singh {\em et al}. Second, they have reported A- values for transitions from the lowest 40 levels alone, and even among these several are missing as may be seen in Table 6. Finally, and most importantly, they have reported A- values only for electric dipole (E1) transitions, whereas in plasma modelling data are also required for the electric quadrupole (E2), magnetic dipole (M1), and magnetic quadrupole (M2) transitions,  as demonstrated by Del Zanna {\em et al}\,  \cite{del04}.  These transitions also affect the subsequent calculations of lifetimes, particularly for those which do not connect with the E1 transitions. Therefore, our {\em aim} is to report a {\em complete} set of  atomic data which can be confidently applied to the modelling of plasmas.

For our calculations we have adopted the {\sc grasp} (general-purpose relativistic atomic structure package) code to generate the wavefunctions. This code was originally developed  by Grant {\em et al} \cite{grasp0} and has been updated by Dr. P. H. Norrington by the name GRASP0. It is fully relativistic and is based on the $jj$ coupling scheme, and is available at the website ({\tt http://web.am.qub.ac.uk/DARC/}). Further relativistic corrections arising from the Breit interaction and QED (quantum electrodynamics) effects have also been included. Like other versions of the {\sc grasp} code, such as {\sc grasp2k} \cite{grasp2k}, this version includes modifications and corrections to the original code, and provides comparable results for both energy levels and radiative rates.  In the calculations, we have used the option of {\em extended average level} (EAL),  in which a weighted (proportional to 2$j$+1) trace of the Hamiltonian matrix is minimised. This produces a compromise set of orbitals describing closely-lying states with moderate accuracy, and generally yields results comparable to other options, such as {\em average level} (AL), as noted by Aggarwal {\em et al}  for several ions of Kr \cite{kr} and Xe \cite{xe}.  

\section{Energy levels}

Although  Ti VII is moderately heavy ($Z$ = 22) and 6 times ionised, CI is  very important for an accurate determination of energy levels. This was also noted earlier for Ti VI \cite{ti6} and Ti X \cite{ti10}. For this reason,  earlier workers have also included CI with additional configurations. For example, Froese-Fischer {\em et al} \cite{cff1} included a very large CI with up to $n \le$ 7 and $\ell \le$ 4 orbitals.  Bearing in mind that our interest is in the lowest $\sim$250 levels and following some tests with a number of configurations, we  have  arrived at the  conclusion that an elaborate CI needs to be included, particularly among the following configurations: 3p$^4$, 3s3p$^5$, 3p$^3$3d, 3p$^6$, 3s3p$^4$3d, 3p$^5$3d, 3p$^2$3d$^2$, 3p$^3$4$\ell$, 3s3p$^3$3d$^2$, 3s3p$^4$4$\ell$,  3p$^2$3d4$\ell$,  3p3d$^3$, 3p$^3$3d4$\ell$,  3p$^4$3d$^2$, 3p$^5$4$\ell$, and 3s3p$^2$3d$^3$. These 31 configurations generate 4186 levels in total and closely interact and intermix. The highest energy range for these levels is up to $\sim$ 15 Ryd and for the lowest 250 levels is below 7 Ryd. Thus the configurations included span a wide energy range.

\subsection{Lowest 48 levels}

In Table 1 we list our calculated energies with the  {\sc grasp} code for the lowest 48 levels of the 3s$^2$3p$^4$, 3s3p$^5$, 3s$^2$3p$^3$3d and 3p$^6$ configurations. Results from calculations with the {\sc grasp} code without (GRASP1) and with (GRASP2) the inclusion of   the Breit and QED corrections are listed. However, the contribution of the Breit and QED effects is negligible. Also included in this table are the experimental energies compiled by NIST \cite{yuri} and the theoretical values obtained by  Bi{\`{e}}mont \cite{eb}, Froese-Fischer {\em et al} \cite{cff1} and Singh {\em et al} \cite{mm4} from the HFR, MCHF and CIV3 codes, respectively. 

All theoretical results are (nearly) consistent in orderings, but the NIST term labels are interchanged for the $^1$D$^o_2$, $^3$P$^o_{0,1,2}$, $^3$D$^o_{1,2,3}$ and $^1$F$^o_3$ levels of the 3p$^3$($^2$P$^o$)3d and 3p$^3$($^2$D$^o$)3d configurations, and hence require a reassessment. We note that these levels are mixed  (see also \cite{mm4}, \cite{eb}), but not highly (except a few), and hence may be unambiguously identified, based on the dominant component of the mixing coefficients, listed in the last column of Table 1. Furthermore, a  similar discrepancy with the NIST listings has also been recently noted for the levels of Ti VI \cite{ti6} and earlier for Kr ions \cite{kr}. Differences in energy levels between our calculated results and the NIST listings (with {\em revised} term labels as listed here) are up to 0.1 Ryd for some of the higher levels. The energies reported by Bi{\`{e}}mont \cite{eb} and Singh {\em et al} \cite{mm4} are closer to the NIST values, because both have made adjustments whereas our calculations are {\em ab initio}. However, energies from the MCHF calculations by Froese-Fischer {\em et al} \cite{cff1} are also closer to the NIST listings, although many levels are missing from the latter.

For levels for which NIST listings are not available, differences between our results from {\sc grasp} and those of Bi{\`{e}}mont \cite{eb} and Singh {\em et al} \cite{mm4}  from the HFR and CIV3 codes are up to 0.15 Ryd for several levels -- see for example levels 10--25 in Table 1, all of which have odd parity. However, the discrepancies, if any,  between the GRASP and MCHF energies are smaller than 0.08 Ryd. This is mainly because both these calculations have included more extensive CI than those of   Bi{\`{e}}mont  and Singh {\em et al}. For example, we have included 15 configurations for generating the odd parity levels whereas Bi{\`{e}}mont  included only 6.  Furthermore, the orderings of some of the $^3$P$^o_{0,1,2}$ and  $^3$D$^o_{1,2,3}$ levels (28--35) in the CIV3 calculations of Singh {\em et al} are inconsistent with our work and that of Froese-Fischer {\em et al} \cite{cff1}.

To further assess the accuracy of our results, we have performed another calculation with the  {\em Flexible Atomic Code} ({\sc fac}) of Gu \cite{fac},  available from the website {\tt {\verb+http://sprg.ssl.berkeley.edu/~mfgu/fac/+}}. This is also a fully relativistic code and provides a variety of atomic parameters with comparable accuracy, particularly  for energy levels and A- values, as already shown for several other ions, see for example:  Aggarwal {\em et al} \cite{fe15} for  Mg-like ions and \cite{ti6}--\cite{ti22} for Ti ions. In addition, an advantage of this code is its high efficiency, which means that reasonably large calculations can be performed within a  time frame of a few weeks. Hence one can experiment with an extensive inclusion of CI in the {\sc  fac} code, and the results obtained are helpful in assessing the accuracy of our energy levels (particularly the higher ones to be discussed later) and radiative rates.

As with the {\sc grasp} code, we have performed a series of calculations using {\sc fac} with increasing amounts of CI. However, here we focus on only three calculations, namely (i) FAC1, which includes exactly the same 4186 levels/configurations as in GRASP, (ii) FAC2, which includes 38,694 levels from all possible combinations of  the 3*6 and 3*5 4*1 configurations, and (iii) FAC3, which also includes levels of the 3*5 5*1 and 3*5 6*1 configurations, i.e. 159,162 in total. This is to date the largest calculation performed by us. The results obtained from these three calculations are also listed in Table 1. All calculations yield (nearly) the same {\em orderings} as in our present results with {\sc grasp},  and agree with each other within 0.04 Ryd. The FAC1 energy levels agree with GRASP2 to within 0.01 Ryd, which means that with the inclusion of the same CI both codes yield comparable results. However, the additional CI  in FAC2 lowers the energies by up to 0.03 Ryd, but  the inclusion of even  larger CI in FAC3 is of no significance, because the two sets of energies (FAC2 and FAC3) agree within 0.01 Ryd.  The agreement between the {\sc grasp} and all {\sc fac} calculations is  within 0.05 Ryd, which is highly satisfactory. Furthermore, the energy levels obtained in FAC2/FAC3  are comparatively in better agreement with the MCHF results of Froese-Fischer {\em et al} \cite{cff1}. This indicates that the inclusion of larger CI (in MCHF and FAC2/FAC3)  improves the accuracy of Ti VII energy levels, and brings theory and experiments closer in agreement. To conclude, we may state that energy levels from the {\sc grasp}, {\sc fac} and MCHF calculations listed in Table 1 are comparatively more accurate than those from the HFR and CIV3 codes. Finally, based on the comparisons made above among a variety of calculations, the energy levels from the {\sc grasp} code are assessed to be accurate  to 0.1 Ryd.

\subsection{3p$^3$4s levels}

The 3p$^3$4s configuration generates 10 levels which are listed in Table 2. NIST energies are available for most of these levels and theoretical results have been reported by  Bi{\`{e}}mont \cite{eb} and Singh {\em et al} \cite{mm4}, which are included in Table 2 for comparison. Our results obtained with the {\sc grasp} and {\sc fac} calculations described above are also included in the table. As for the levels of Table 1, the Breit and QED contributions are negligible and there is no discrepancy between the GRASP2 and FAC1 energies.  Those  obtained by Bi{\`{e}}mont  and Singh {\em et al} agree with each other as well as with NIST, because of the adjustments made, whereas our results from {\sc grasp} are higher by up to 0.1 Ryd.  However, the additional CI included in FAC2 lowers the energies by up to 0.1 Ryd and brings the results closer to the experimental values {\em without} any adjustment. As with the levels of Table 1, further inclusion of CI in FAC3 is of no additional advantage.

\subsection{3s3p$^4$3d levels}

The 3s3p$^4$3d configuration generates 56 levels, listed in Table 3. The only other results available in the literature are from the recent calculations of   Singh {\em et al} \cite{mm4} obtained with the CIV3 code. Their energy levels are listed in Table 3 along with our calculations with the {\sc grasp} and {\sc fac} codes. As with the levels of Tables 1 and 2, the additional CI included in the FAC3 calculations is of no advantage over that included in FAC2, because both sets of energies agree within 0.01 Ryd for all levels. Energies for most of the levels from our calculations with {\sc grasp} are also in close agreement with FAC1, and the differences with FAC2 for a few (such as 39--47) are smaller than 0.05 Ryd. More importantly, both codes yield the same orderings, whereas that in the CIV3 calculation is different for a few, such as 18--20 and 26--27. Furthermore, Singh {\em et al} have interchanged the $^3$P$^o_{0,1,2}$ (14--16 and 45--47)  levels of the 3s3p$^4$($^1$D)3d and 3s3p$^4$($^3$P)3d configurations, and have incorrectly identified level 34 as  3s3p$^4$($^1$S)3d  $^1$F$_3$ instead of 3s3p$^4$($^1$D)3d $^1$F$_3$. The CIV3 energies calculated by Singh {\em et al} are the {\em highest} among all the results listed in Table 3, and differ by up to 1 Ryd for some of the levels -- see for example, 30--34 and particularly 35--37. Their overestimation of energies  is clearly due to the inclusion of inadequate CI, as has also been noted  for the levels of Table 1. 

\subsection{Lowest 231 levels}

In Table 4 (see supplementary data, available online at stacks.iop.org/PhysScr/vol/number/mmedia) we list our final energies, in increasing order,  obtained using the {\sc grasp} code with CI among 31 configurations listed in  section 2, which correspond to the GRASP2 calculations. These configurations generate 4186 levels, but for conciseness energies are listed only for the lowest 231 levels, which include all those of the 3p$^4$, 3s3p$^5$, 3p$^3$3d, 3p$^6$, 3s3p$^4$3d,  3p$^3$4s and 3p$^3$4p  configurations, but only a few of the others, such as  3p$^2$3d$^2$. However, energies corresponding to any of the calculations described in section 2.1 and for any  number of levels up to 159,162 can be obtained on request from the first author (K.Aggarwal@qub.ac.uk).

Although calculations with the {\sc fac} code have been performed with the inclusion of larger CI, energies obtained with the {\sc grasp} code alone are listed in Table 4 (see supplementary data, available online at stacks.iop.org/PhysScr/vol/number/mmedia). This is partly because both codes provide energies with comparable accuracy as demonstrated and discussed in sections 2.1 to 2.3, but mainly because the $LSJ$ designations of the levels are also determined  in the {\sc grasp} code. For a majority of users these designations are more familiar and hence preferable. However, we note that the $LSJ$ designations provided in this table are not always unique, because some of the levels are highly mixed, mostly from the same but sometimes with other configurations. Therefore as in Table 1, in Table 4 also we have listed mixing coefficients. This problem is common to most calculations, particularly for those ions in which CI is very important, such as Ti VI \cite{ti6} and Ti X \cite{ti10}. Furthermore, because of mixing it is easier to identify levels in a GRASP calculation than in FAC.  Therefore, care has been taken to provide the most appropriate designation of a level/configuration, but a redesignation of  these cannot be ruled out in a few cases, such as for levels 86 (3p$^2$($^3$P)3d$^2$($^3$P)   $^3$P$  _1$), 141 (3s3p$^4$($^1$D)3d   $^1$S$  _0$) and 176 (3s3p$^4$($^3$P)3d  $^1$D$  _2$). 

For the 231 levels listed in Table 4 (see supplementary data, available online at stacks.iop.org/PhysScr/vol/number/mmedia), comparisons with the NIST compilations of experimental energies has been possible for only a few, as shown in Tables 1 and 2. There are no major discrepancies  with our calculations, although the term labels  of the levels differ in a few instances as noted in section 2.1. However, extensive comparisons have been possible, for about half of the levels, with  other available theoretical work, particularly of  Singh {\em et al} \cite{mm4},  as  shown in Tables 1--3. Based on these comparisons it is concluded that CI is very important for the energy levels of Ti VII, but mostly among those configurations whose levels interact closely.  Singh {\em et al} also included a large CI in their calculations with the CIV3 code \cite{civ3}, but that is not as extensive as considered in the present work or in the earlier calculations of Froese-Fischer {\em et al} \cite{cff1}.  Similarly, they adjusted their calculated energies using the NIST compilations, but this has not been useful as  experimental energies are  available for only a few. For these reasons, differences between our calculations and those reported by Singh {\em et al} are significant (up to 1 Ryd) for many levels, and their orderings also differ in a few instances. On the other hand, our GRASP2 and FAC1  energies are comparable for most of the levels, in both magnitude as well as orderings. Thus we have confidence in our results, and based on a variety of comparisons assess the accuracy of our energy levels listed in Table 4 (see supplementary data, available online at stacks.iop.org/PhysScr/vol/number/mmedia)  to be better than 1\%. 

\section{Radiative rates}

The absorption oscillator strength ($f_{ij}$) and radiative rate A$_{ji}$ (in s$^{-1}$) for a transition $i \to j$ are related by the following expression:

\begin{equation}
f_{ij} = \frac{mc}{8{\pi}^2{e^2}}{\lambda^2_{ji}} \frac{{\omega}_j}{{\omega}_i}A_{ji}
 = 1.49 \times 10^{-16} \lambda^2_{ji} (\omega_j/\omega_i) A_{ji}
\end{equation}
where $m$ and $e$ are the electron mass and charge, respectively, $c$  the velocity of light, 
$\lambda_{ji}$  the transition energy/wavelength in $\rm \AA$, and $\omega_i$ and $\omega_j$ the statistical weights of the lower $i$ and upper $j$ levels, respectively.
Similarly, the oscillator strength $f_{ij}$ (dimensionless) and the line strength $S$ (in atomic unit, 1 a.u. = 6.460$\times$10$^{-36}$ cm$^2$ esu$^2$) are related by the 
following standard equations:

\begin{flushleft}
for the electric dipole (E1) transitions: 
\end{flushleft} 
\begin{equation}
A_{ji} = \frac{2.0261\times{10^{18}}}{{{\omega}_j}\lambda^3_{ji}} S \hspace*{1.0 cm} {\rm and} \hspace*{1.0 cm} 
f_{ij} = \frac{303.75}{\lambda_{ji}\omega_i} S, \\
\end{equation}
\begin{flushleft}
for the magnetic dipole (M1) transitions:  
\end{flushleft}
\begin{equation}
A_{ji} = \frac{2.6974\times{10^{13}}}{{{\omega}_j}\lambda^3_{ji}} S \hspace*{1.0 cm} {\rm and} \hspace*{1.0 cm}
f_{ij} = \frac{4.044\times{10^{-3}}}{\lambda_{ji}\omega_i} S, \\
\end{equation}
\begin{flushleft}
for the electric quadrupole (E2) transitions: 
\end{flushleft}
\begin{equation}
A_{ji} = \frac{1.1199\times{10^{18}}}{{{\omega}_j}\lambda^5_{ji}} S \hspace*{1.0 cm} {\rm and} \hspace*{1.0 cm}
f_{ij} = \frac{167.89}{\lambda^3_{ji}\omega_i} S, 
\end{equation}

\begin{flushleft}
and for the magnetic quadrupole (M2) transitions: 
\end{flushleft}
\begin{equation}
A_{ji} = \frac{1.4910\times{10^{13}}}{{{\omega}_j}\lambda^5_{ji}} S \hspace*{1.0 cm} {\rm and} \hspace*{1.0 cm}
f_{ij} = \frac{2.236\times{10^{-3}}}{\lambda^3_{ji}\omega_i} S. \\
\end{equation}

The A- and f- values have been calculated in both Babushkin and Coulomb gauges, which are  equivalent to the length and velocity forms in the non-relativistic nomenclature. However, the results are presented here in the length form alone  which are considered to be comparatively more accurate  \cite{ipg} -- \cite{jps}.   In Table 5 (see supplementary data, available online at stacks.iop.org/PhysScr/vol/number/mmedia) we present transition energies ($\Delta$E$_{ij}$ in ${\rm \AA}$), radiative rates (A$_{ji}$ in s$^{-1}$), oscillator strengths ($f_{ij}$, dimensionless), and line strengths ($S$ in a.u.) for all 5758 electric dipole (E1) transitions among the lowest 231 levels of Ti VII. The {\em indices} used to represent the lower and upper levels of a transition have already been defined in Table 4 (see supplementary data, available online at stacks.iop.org/PhysScr/vol/number/mmedia). Also, in calculating the  above parameters we have used the Breit and QED-corrected theoretical energies/wavelengths as listed in Table 4. However, only A- values are included in Table 5 for the 11,111 electric quadrupole (E2), 7994  magnetic dipole (M1), and 7870 magnetic quadrupole (M2) transitions. Corresponding results for f- or S- values can be easily obtained by using Eqs. (1-5). 

In Table 6  we compare our f- values for transitions from the levels of 3p$^4$ to the 3s3p$^5$ and 3p$^3$3d configurations of Ti VII, which are {\em common} to all existing calculations. It may be noted that level indices representing transitions correspond  to those of Table 4 available online at stacks.iop.org/PhysScr/vol/number/mmedia. Included in this table are our results from  {\sc grasp} and {\sc fac} (FAC1),  plus the earlier calculations of Froese-Fischer {\em et al} \cite{cff1}, Singh {\em et al} \cite{mm4}  and Bi{\`{e}}mont \cite{eb} from the  MCHF, CIV3  and HFR codes, respectively.  It is highly satisfactory to note that both calculations from {\sc grasp} and {\sc fac} provide comparable f- values for almost all transitions,  both strong as well as weak. Differences for a few transitions between the two calculations (such as 1--16 and 2--15) are within a factor of three. The only transition for which the two results disagree by two orders of magnitude is 2--34 (3p$^4$ $^3$P$_1$ -- 3p$^3$($^2$P$^o$)3d $^3$D$^o_2$) with f=6$\times$10$^{-4}$ in GRASP and f=5$\times$10$^{-6}$ in FAC. Therefore, we can say with confidence that our calculated results of radiative rates are reliable.

Many transitions are unfortunately missing from the CIV3 calculations of Singh {\em et al} \cite{mm4} and some of these are comparatively strong, such as: 1--41, 2--41/45 and 3--41. Furthermore, discrepancies between the CIV3 f- values and other calculations listed in Table 6 are up to five orders of magnitude -- see for example, transitions 1--16/31/33, 2--15/31/33/34, 3--17/31 and 4--36. In some cases the CIV3 f- values are larger (such as 1--16) whereas for others  they are smaller (such as 1--31), and hence there is no systematic trend in the discrepancies. From the comparisons shown in Table 6  it can be concluded with confidence that not only are the results of Singh {\em et al} incomplete but  also unreliable. The main reason, in our opinion, for these large discrepancies is the inadequate inclusion of CI in their calculations. Furthermore, a normal practice in a CIV3 calculation is to first survey all levels of a configuration and then eliminate those whose eigenvectors are below a certain magnitude (say $\sim$ 0.2) before performing a final run for transition rates. This exercise is undertaken to keep the calculations manageable within the limited computational resources available.  This elimination process particularly affects the weak(er) transitions, say with f $\le$ 0.001.  Similar differences, and for the same reasons, were noted by Aggarwal {\em et al} \cite{feix} in their calculations for transitions in Fe IX  \cite{mm3}, and more recently by Aggarwal and Keenan for  Ti VI \cite{ti6} and Ti X \cite{ti10}. We also note that (perhaps) Singh {\em et al} have relied too much on improving the accuracy of their energy levels by making adjustments based on the NIST listings. As discussed  earlier \cite{fst}, we emphasise  that the process of fine-tuning may make the theoretical energy levels more accurate in magnitude, but {\em not} the subsequent calculations of f- values (or other parameters such as lifetimes and collision strengths), if inherent limitations are already present. 

The HFR f- values of Bi{\`{e}}mont \cite{eb} are comparable with our results with {\sc grasp} or the earlier ones of  Froese-Fischer {\em et al} \cite{cff1} with the MCHF code for most of the strong transitions, but there are discrepancies for the weaker ones of up to three orders of magnitude for a few, such as: 1--31, 2--34 and 3--17. The main reason for these large discrepancies is the insufficient inclusion of CI in the calculations of  Bi{\`{e}}mont. However, generally there is a good agreement between our f- values and those of Froese-Fischer {\em et al} for most of the transitions, strong as well as weak. This is primarily because both calculations have included extensive CI in the generation of transition rates. Nevertheless, there are also a few transitions (such as 1--16/41, 2--26/32/34 and 4--31/32) for which the differences are noticeable, but are still within an order of magnitude. Considering that many of the transitions listed in Table 6 have very small f- values, this general agreement among three independent calculations by three different methods is highly satisfactory, and confirms that our results listed in Table 5 (see supplementary data, available online at stacks.iop.org/PhysScr/vol/number/mmedia) are accurate to $\sim$20\% for a majority of transitions, particularly the strong ones with f  $>$ 0.001. 

One of the general criteria to assess the accuracy of radiative rates is to compare the length and velocity forms of the f- or A- values, although the primary parameter to calculate is the S- value. This is because A- values are (normally) employed in the modelling of plasmas (along with other parameters), but are  comparatively large in magnitude. On the other hand f- values are smaller, easy on the eyes, and give an indication of the strength of a transition. Furthermore, for $\sim$ 60\% of the E1 transitions f- values have magnitudes similar to the S- values (see Table 5), i.e. within an order of magnitude,  and therefore make a sense to compare. However, such comparisons are flexible and only desirable, but are {\em not} a fully sufficient test to assess accuracy, as calculations based on different methods (or combinations of configurations) may give comparable f- values in the two forms, but entirely different results in magnitude. Nevertheless, in Table 5 we have also listed the velocity/length (i.e. Coulomb/Babushkin) ratio of the A- values, which are directly proportional to the S- values as seen in Eq. (2). Generally, there is good agreement between the length and velocity forms of the f- values for {\em strong} transitions (f $\ge$ 0.01) as already seen in Table 6, but differences  between the two can sometimes be substantial, even for some very strong transitions, as demonstrated by several  examples by Aggarwal {\em et al} \cite{fe15}. Nevertheless, for almost all of the strong E1 transitions  the two forms agree to within 20\%, but the differences for 163 ($<$3\%) of the transitions are slightly larger. In fact, for only 14 transitions do the f- values differ by over  50\%, but are still within a factor of three. Therefore, on the basis of these and earlier comparisons shown in Table 6 we may reaffirm  that for a majority of the strong E1 transitions, our radiative rates are accurate to better than 20\%. However, for the weaker transitions this assessment of accuracy does not apply, because such transitions are very sensitive to mixing coefficients, and hence differing amount of CI (and methods) produce different A- values, as discussed in detail by Hibbert \cite{ah3}. This is the main reason that the two forms of f- values for some weak transitions sometimes differ significantly (by orders of magnitude), and examples include 2--214 (f  = 7.2$\times$10$^{-15}$), 9--66 (f = 1.6$\times$10$^{-7}$) and 21--49 (f = 6.7$\times$10$^{-6}$). Although f- values for weak transitions may be required in plasma modelling for completeness,  their contribution is normally less important compared to stronger transitions with f $\ge$ 0.001. For this reason many authors (and some codes) do not normally report A- values for very weak transitions.

The accuracy assessment made above for the f- or A- values of E1 transitions is mostly based on the comparisons made in Table 6 and the ratio of their velocity and length forms discussed above, but are mostly applicable for  transitions with significant magnitude of f- or S- values.  Any other criteria or comparison with other calculations with comparable complexity, preferably by another method/code, may lead to a (slightly) different conclusion. A similar comparison for the E2, M1 and M2 transitions is not possible, mainly because they are comparatively much weaker, and hence susceptible to vary with differing amounts of CI. Furthermore, there are no similar data available in the literature with which to make comparisons.
 
\section{Lifetimes}

The lifetime $\tau$ of a level $j$ is defined as follows:

\begin{equation}
{\tau}_j = \frac{1}{{\sum_{i}^{}} A_{ji}}.
\end{equation}

In Table 4 (see supplementary data, available online at stacks.iop.org/PhysScr/vol/number/mmedia)  we list lifetimes for all 231 levels from our calculations with the {\sc grasp} code (corresponding to GRASP2). These results {\em include} A- values from all types of transitions, i.e. E1, E2, M1 and M2. 

Lifetimes for two levels, i.e.  3s3p$^5$ $^3$P$^0_2$ (6) and 3s3p$^5$ $^1$P$^0_1$ (9), have been measured by Dumont {\em et al} \cite{pdd}  to be 0.66$\pm$0.05 and 0.30$\pm$0.02 ns, respectively. For both levels, E1 transitions dominate and the measurements agree well with our theoretical results of  0.72 and 0.24 ns, respectively. Measurements of lifetimes for additional levels of Ti VII would be helpful for a further assessment of the accuracy of our calculations.

\section{Conclusions}

In the present work, energy levels, radiative rates, oscillator strengths and line strengths for transitions among 231 fine-structure levels of Ti VII are computed  using the fully relativistic {\sc grasp} code, and results reported for electric and magnetic dipole and quadrupole transitions. For calculating these parameters an extensive CI (with up to 4186 levels) has been included, which has been observed to be very significant, particularly for the accurate determination of energy levels. Furthermore, analogous calculations have been performed with the {\sc fac} code and with the inclusion of even larger CI with up to 159,162 levels, but most of the additional configurations included do not appreciably affect the magnitude or orderings of the lowest 231 energy levels considered in this work. Based on a variety of comparisons among different calculations, the reported energy levels are assessed to be accurate to better than 1\%. 

There is a paucity of measured energies for a majority of  the levels of Ti VII. However, for the common levels there is no major discrepancy  with our calculations, although the orderings slightly differ in a few instances. Other theoretical energies are available, mostly from a recent calculation by Singh {\em et al} \cite{mm4}, but only for about half the levels. However, discrepancies with their results are up to 1 Ryd, and based on several comparisons  their listed energies are not assessed to be accurate.  Discrepancies are even greater, up to five orders of magnitude,  for the f- values between their data and the present as well as earlier calculations. As for the energy levels, extensive comparisons, based on a variety of calculations with the {\sc grasp} and {\sc fac} codes,  have been made for the f- values, and the accuracy of these is assessed to be  $\sim$ 20\% for a majority of the strong transitions. 

Lifetimes are also reported for all levels, but  measurements are available for only two, for which there is a good agreement with theory. Finally, calculations for energy levels  have been performed for up to 159,162 levels of Ti VII, and for radiative rates up to 4186 levels, but for brevity results have been reported for only the lowest 231 levels. However, a complete set of results for all calculated parameters can be obtained on request from one of the authors (K.Aggarwal@qub.ac.uk).

\section*{Acknowledgment}
KMA  is thankful to  AWE Aldermaston for financial support.   



\clearpage
\newpage
\begin{flushleft}
{\bf Table 1.} Lowest 48 levels  of Ti VII and their excitation energies (in Ryd).
\newline 
\end{flushleft}
{\tiny
\begin{tabular}{rlllrrrrrrrrl} \hline
Index    & Configuration            & Level              &   NIST    &  GRASP1	& GRASP2  &  FAC1   &  FAC2  & FAC3   &  HFR    &  MCHF   & CIV3    & MC \\
\hline
   1  &  3s$^2$3p$^4$		    &  $^3$P$   _2$      &  0.00000  &  0.0000  & 0.0000  & 0.0000  & 0.0000 & 0.0000 &  0.0000 & 0.0000  &  0.0000 &	0.972  \\  
   2  &  3s$^2$3p$^4$	      	    &  $^3$P$   _1$      &  0.04132  &  0.0416  & 0.0405  & 0.0403  & 0.0402 & 0.0403 &  0.0413 & 0.0400  &  0.0417 &	0.979  \\ 
   3  &  3s$^2$3p$^4$	      	    &  $^3$P$   _0$      &  0.05366  &  0.0545  & 0.0531  & 0.0529  & 0.0527 & 0.0528 &  0.0538 & 0.0515  &  0.0554 &  -0.972  \\ 
   4  &  3s$^2$3p$^4$	      	    &  $^1$D$   _2$      &  0.21989  &  0.2397  & 0.2387  & 0.2362  & 0.2361 & 0.2348 &  0.2196 & 0.2261  &  0.2129 &  -0.971  \\ 
   5  &  3s$^2$3p$^4$	      	    &  $^1$S$   _0$      &  0.49938  &  0.5227  & 0.5217  & 0.5157  & 0.5152 & 0.5164 &  0.4992 & 0.5019  &  0.4996 &	0.959  \\ 
   6  &  3s3p$^5$	      	    &  $^3$P$^o _2$      &  1.78851  &  1.7812  & 1.7795  & 1.7815  & 1.7737 & 1.7753 &  1.7916 & 1.7733  &  1.7822 &	0.879  \\ 
   7  &  3s3p$^5$	      	    &  $^3$P$^o _1$      &  1.82307  &  1.8163  & 1.8137  & 1.8155  & 1.8076 & 1.8092 &  1.8230 & 1.8064  &  1.8229 &	0.877  \\ 
   8  &  3s3p$^5$	      	    &  $^3$P$^o _0$      &  1.84260  &  1.8358  & 1.8328  & 1.8345  & 1.8266 & 1.8282 &  1.8409 & 1.8248  &  1.8443 &  -0.878  \\ 
   9  &  3s3p$^5$	      	    &  $^1$P$^o _1$      &  2.28923  &  2.3309  & 2.3280  & 2.3260  & 2.3074 & 2.3046 &  2.2881 & 2.2844  &  2.2829 &	0.729+0.622(41)  \\ 
  10  &  3s$^2$3p$^3$($^4$S$^o$)3d  &  $^5$D$^o _0$      &  	     &  2.4659  & 2.4626  & 2.4628  & 2.4348 & 2.4319 &  2.3556 & 2.4444  &  2.3803 &  -0.983  \\ 
  11  &  3s$^2$3p$^3$($^4$S$^o$)3d  &  $^5$D$^o _1$      &  	     &  2.4664  & 2.4630  & 2.4632  & 2.4352 & 2.4323 &  2.3561 & 2.4450  &  2.3803 &	0.983  \\ 
  12  &  3s$^2$3p$^3$($^4$S$^o$)3d  &  $^5$D$^o _2$      &  	     &  2.4674  & 2.4637  & 2.4639  & 2.4359 & 2.4329 &  2.3570 & 2.4460  &  2.3803 &  -0.982  \\ 
  13  &  3s$^2$3p$^3$($^4$S$^o$)3d  &  $^5$D$^o _3$      &  	     &  2.4688  & 2.4647  & 2.4649  & 2.4369 & 2.4339 &  2.3583 & 2.4474  &  2.3804 &  -0.981  \\ 
  14  &  3s$^2$3p$^3$($^4$S$^o$)3d  &  $^5$D$^o _4$      &  	     &  2.4711  & 2.4666  & 2.4668  & 2.4387 & 2.4357 &  2.3605 &	  &  2.3804 &	0.982  \\ 
  15  &  3s$^2$3p$^3$($^2$D$^o$)3d  &  $^3$D$^o _2$      &  	     &  2.6682  & 2.6643  & 2.6619  & 2.6313 & 2.6270 &  2.5449 & 2.6257  &  2.5009 &  -0.691-0.629(43)  \\ 
  16  &  3s$^2$3p$^3$($^2$D$^o$)3d  &  $^3$D$^o _3$      &  	     &  2.6704  & 2.6662  & 2.6637  & 2.6331 & 2.6288 &  2.5456 & 2.6279  &  2.5094 &	0.702+0.632(42)  \\ 
  17  &  3s$^2$3p$^3$($^2$D$^o$)3d  &  $^3$D$^o _1$      &  	     &  2.6757  & 2.6715  & 2.6690  & 2.6383 & 2.6340 &  2.5487 & 2.6329  &  2.5197 &	0.708+0.644(44)  \\ 
  18  &  3s$^2$3p$^3$($^2$D$^o$)3d  &  $^3$F$^o _2$      &  	     &  2.7389  & 2.7356  & 2.7326  & 2.7033 & 2.6983 &  2.6466 & 2.6924  &  2.6708 &	0.854  \\ 
  19  &  3s$^2$3p$^3$($^2$D$^o$)3d  &  $^1$S$^o _0$      &  	     &  2.7405  & 2.7364  & 2.7342  & 2.6988 & 2.6938 &  2.6202 & 2.6974  &  2.6459 &	0.977  \\ 
  20  &  3s$^2$3p$^3$($^2$D$^o$)3d  &  $^3$F$^o _3$      &  	     &  2.7524  & 2.7482  & 2.7451  & 2.7156 & 2.7106 &  2.6554 & 2.7047  &  2.6801 &	0.866  \\ 
  21  &  3s$^2$3p$^3$($^2$D$^o$)3d  &  $^3$F$^o _4$      &  	     &  2.7689  & 2.7637  & 2.7604  & 2.7309 & 2.7258 &  2.6669 &	  &  2.6920 &	0.882  \\ 
  22  &  3s$^2$3p$^3$($^2$D$^o$)3d  &  $^3$G$^o _3$      &  	     &  2.9345  & 2.9308  & 2.9248  & 2.8954 & 2.8886 &  2.7980 & 2.8690  &  2.8667 &	0.975  \\ 
  23  &  3s$^2$3p$^3$($^2$D$^o$)3d  &  $^3$G$^o _4$      &  	     &  2.9383  & 2.9341  & 2.9280  & 2.8986 & 2.8917 &  2.8010 &	  &  2.8667 &	0.973  \\ 
  24  &  3s$^2$3p$^3$($^2$D$^o$)3d  &  $^3$G$^o _5$      &  	     &  2.9441  & 2.9389  & 2.9327  & 2.9034 & 2.8965 &  2.8048 &	  &  2.8667 &	0.985  \\ 
  25  &  3s$^2$3p$^3$($^2$D$^o$)3d  &  $^1$G$^o _4$      &  	     &  3.0106  & 3.0063  & 2.9987  & 2.9688 & 2.9616 &  2.8521 &	  &  2.9303 &	0.966  \\ 
  26  &  3s$^2$3p$^3$($^2$P$^o$)3d  &  $^1$D$^o _2$      &  2.96400  &  3.0294  & 3.0251  & 3.0188  & 2.9952 & 2.9901 &  2.9666 & 2.9691  &  2.9627 &	0.863  \\ 
  27  &  3s$^2$3p$^3$($^2$P$^o$)3d  &  $^3$F$^o _4$      &  	     &  3.1511  & 3.1460  & 3.1399  & 3.1158 & 3.1098 &  3.1187 &	  &  3.0972 &	0.863  \\ 
  28  &  3s$^2$3p$^3$($^2$P$^o$)3d  &  $^3$P$^o _0$      &           &  3.1536  & 3.1507  & 3.1448  & 3.1198 & 3.1150 &  3.0949 & 3.0960  &  3.0069 &  -0.861  \\ 
  29  &  3s$^2$3p$^3$($^2$P$^o$)3d  &  $^3$F$^o _3$      &  	     &  3.1539  & 3.1493  & 3.1432  & 3.1191 & 3.1132 &  3.1213 & 3.0893  &  3.1010 &	0.863  \\ 
  30  &  3s$^2$3p$^3$($^2$P$^o$)3d  &  $^3$F$^o _2$      &  	     &  3.1607  & 3.1564  & 3.1502  & 3.1260 & 3.1200 &  3.1259 & 3.0958  &  3.1192 &	0.847  \\ 
  31  &  3s$^2$3p$^3$($^2$P$^o$)3d  &  $^3$P$^o _1$      &           &  3.1648  & 3.1613  & 3.1553  & 3.1304 & 3.1256 &  3.1070 & 3.1075  &  3.0243 &	0.859  \\ 
  32  &  3s$^2$3p$^3$($^2$P$^o$)3d  &  $^3$D$^o _1$      &           &  3.1716  & 3.1688  & 3.1612  & 3.1361 & 3.1308 &  3.0622 & 3.1043  &  3.0113 &	0.748  \\ 
  33  &  3s$^2$3p$^3$($^2$P$^o$)3d  &  $^3$P$^o _2$      &           &  3.1932  & 3.1887  & 3.1820  & 3.1574 & 3.1524 &  3.1358 & 3.1366  &  3.1056 &	0.704  \\ 
  34  &  3s$^2$3p$^3$($^2$P$^o$)3d  &  $^3$D$^o _2$      &           &  3.1974  & 3.1932  & 3.1859  & 3.1609 & 3.1558 &  3.0814 & 3.1274  &  3.0516 &	0.586+0.540(33)+0.428(15)  \\ 
  35  &  3s$^2$3p$^3$($^2$P$^o$)3d  &  $^3$D$^o _3$      &           &  3.2270  & 3.2218  & 3.2139  & 3.1888 & 3.1835 &  3.1066 & 3.1572  &  3.1142 &	0.760  \\ 
  36  &  3s$^2$3p$^3$($^2$P$^o$)3d  &  $^1$F$^o _3$      &  3.29791  &  3.4407  & 3.4361  & 3.4249  & 3.3992 & 3.3929 &  3.2971 & 3.3562  &  3.2985 &	0.801  \\ 
  37  &  3s$^2$3p$^3$($^2$D$^o$)3d  &  $^3$S$^o _1$      &  3.41939  &  3.5370  & 3.5329  & 3.5225  & 3.4991 & 3.4880 &  3.4099 & 3.4441  &  3.4197 &  -0.937  \\ 
  38  &  3s$^2$3p$^3$($^2$D$^o$)3d  &  $^3$P$^o _2$      &  3.44107  &  3.5455  & 3.5418  & 3.5274  & 3.4954 & 3.4915 &  3.4403 & 3.4795  &  3.4443 &	0.825  \\ 
  39  &  3s$^2$3p$^3$($^2$D$^o$)3d  &  $^3$P$^o _1$      &  3.45254  &  3.5628  & 3.5590  & 3.5441  & 3.5135 & 3.5089 &  3.4536 & 3.4931  &  3.4521 &  -0.631-0.408(41)  \\ 
  40  &  3s$^2$3p$^3$($^2$D$^o$)3d  &  $^3$P$^o _0$      &  3.47929  &  3.5842  & 3.5796  & 3.5652  & 3.5335 & 3.5297 &  3.4801 & 3.5163  &  3.4844 &	0.790+0.430(8)  \\ 
  41  &  3s$^2$3p$^3$($^2$D$^o$)3d  &  $^1$P$^o _1$      &  3.48007  &  3.5932  & 3.5889  & 3.5728  & 3.5405 & 3.5373 &  3.4828 & 3.5208  &  3.5024 &  -0.612+0.512(9)  \\ 
  42  &  3s$^2$3p$^3$($^4$S$^o$)3d  &  $^3$D$^o _3$      &  3.58736  &  3.7076  & 3.7035  & 3.6901  & 3.6595 & 3.6510 &  3.5904 & 3.6244  &  3.6097 &	0.674+0.545(35)  \\ 
  43  &  3s$^2$3p$^3$($^4$S$^o$)3d  &  $^3$D$^o _2$      &  3.61383  &  3.7344  & 3.7298  & 3.7162  & 3.6857 & 3.6772 &  3.6135 & 3.6502  &  3.6226 &  -0.661-0.555(34)  \\ 
  44  &  3s$^2$3p$^3$($^4$S$^o$)3d  &  $^3$D$^o _1$      &  3.63165  &  3.7521  & 3.7470  & 3.7334  & 3.7029 & 3.6944 &  3.6292 & 3.6674  &  3.6383 &  -0.659-0.570(32)  \\ 
  45  &  3s$^2$3p$^3$($^2$D$^o$)3d  &  $^1$D$^o _2$      &  3.71526  &  3.8551  & 3.8504  & 3.8365  & 3.8100 & 3.7976 &  3.7250 & 3.7553  &  3.7188 &	0.867  \\ 
  46  &  3p$^6$ 		    &  $^1$S$   _0$      &  	     &  3.9405  & 3.9367  & 3.9426  & 3.9270 & 3.9290 & 	&	  &  3.9669 &	0.710  \\ 
  47  &  3s$^2$3p$^3$($^2$D$^o$)3d  &  $^1$F$^o _3$      &  3.83208  &  3.9742  & 3.9693  & 3.9521  & 3.9161 & 3.9081 &  3.8347 &	  &  3.8780 &	0.800  \\ 
  48  &  3s$^2$3p$^3$($^2$P$^o$)3d  &  $^1$P$^o _1$      &  4.10734  &  4.1488  & 4.1442  & 4.1280  & 4.1076 & 4.0986 &  4.1022 & 4.0718  &  4.1010 &	0.936  \\ 
 \hline				  								   	   
\end{tabular}	
}												   		   
\begin {flushleft}														   
\begin{tabbing}
aaaaaaaaaaaaaaaaaaaaaaaaaaaaaaaaaaaa\= \kill
NIST:  \cite{yuri}, for some of the levels (26 and higher) listed  designations are {\em different}, but have been revised here\\ -- see the text in section 2.1 \\ 
GRASP1: Present results {\em without} QED effects   \\
GRASP2: Present results {\em with} QED effects \\
FAC1: Present results with 4186 levels \\
FAC2: Present results with 38,694 levels  \\
FAC3: Present results with 159,162 levels  \\
HFR: Results of Biemont    \cite{eb} \\ 
MCHF: Results of Froese-Fischer {\em et al}  \cite{cff1} \\ 
CIV3: Results of Singh {\em et al} \cite{mm4} \\ 
MC: Mixing coefficients for the {\sc grasp} calculations, the first number corresponds to the designated level \\and subsequent one(s) to the level(s) inside the bracket \\
\end{tabbing}
\end {flushleft}

\clearpage
\newpage
\begin{flushleft}
{\bf Table 2.} Levels  of the 3s$^2$3p$^3$4s configuration of Ti VII and their excitation energies (in Ryd).
\newline 
\end{flushleft}
{\small 
\begin{tabular}{rllrrrrrrrrrr} \hline
Index    & Configuration & Level             &   NIST  &  GRASP1 & GRASP2 &  FAC1   &  FAC2  & FAC3    &  HFR   &  CIV3  \\
\hline
    1  & 3s$^2$3p$^3$4s  &   $^5$S$^o _ 2$   &         & 5.0305  & 5.0266 & 5.03764  & 4.9801 & 4.9781 & 5.0255 & 4.9979 \\ 
    2  & 3s$^2$3p$^3$4s  &   $^3$S$^o _ 1$   & 5.14152 & 5.1651  & 5.1613 & 5.17865  & 5.1212 & 5.1221 & 5.1416 & 5.1444 \\ 
    3  & 3s$^2$3p$^3$4s  &   $^3$D$^o _ 1$   & 5.34086 & 5.4039  & 5.4005 & 5.40127  & 5.3276 & 5.3259 & 5.3410 & 5.3415 \\ 
    4  & 3s$^2$3p$^3$4s  &   $^3$D$^o _ 2$   & 5.34283 & 5.4060  & 5.4024 & 5.40274  & 5.3292 & 5.3275 & 5.3430 & 5.3427 \\ 
    5  & 3s$^2$3p$^3$4s  &   $^3$D$^o _ 3$   & 5.34912 & 5.4117  & 5.4074 & 5.40727  & 5.3348 & 5.3331 & 5.3495 & 5.3497 \\ 
    6  & 3s$^2$3p$^3$4s  &   $^1$D$^o _ 2$   & 5.40307 & 5.4741  & 5.4703 & 5.47346  & 5.3999 & 5.3995 & 5.4028 & 5.4046 \\ 
    7  & 3s$^2$3p$^3$4s  &   $^3$P$^o _ 0$   & 5.53629 & 5.6443  & 5.6406 & 5.63255  & 5.5208 & 5.5188 & 5.5364 & 5.5371 \\ 
    8  & 3s$^2$3p$^3$4s  &   $^3$P$^o _ 1$   & 5.54034 & 5.6479  & 5.6440 & 5.63652  & 5.5253 & 5.5232 & 5.5404 & 5.5448 \\ 
    9  & 3s$^2$3p$^3$4s  &   $^3$P$^o _ 2$   & 5.55067 & 5.6566  & 5.6521 & 5.64623  & 5.5367 & 5.5347 & 5.5505 & 5.5502 \\ 
   10  & 3s$^2$3p$^3$4s  &   $^1$P$^o _ 1$   & 5.60242 & 5.7172  & 5.7129 & 5.70953  & 5.5989 & 5.5980 & 5.6029 & 5.6086 \\ 
 \hline				  								   	   
\end{tabular}	
}												   		   
\begin {flushleft}														   
\begin{tabbing}
aaaaaaaaaaaaaaaaaaaaaaaaaaaaaaaaaaaa\= \kill
NIST:  \cite{yuri} \\
GRASP1: Present results {\em without} QED effects   \\
GRASP2: Present results {\em with} QED effects \\
FAC1: Present results with  4186 levels \\
FAC2: Present results with 38,694 levels \\
FAC3: Present results with 159,162 levels \\
HFR: Results of Biemont  \cite{eb} \\ 
CIV3: Results of Singh {\em et al}   \cite{mm4}  \\ 
\end{tabbing}
\end {flushleft}

\clearpage
\newpage
\begin{flushleft}
Table 3. Levels of the 3s3p$^4$3d configuration of Ti VII and their excitation energies (in Ryd). 
\end{flushleft}
\begin{tabular}{rllrlrllrl} \hline
Index  & \multicolumn{2}{c}{Configuration/Level} &                   GRASP2  & FAC1      & FAC2   & FAC3     & CIV3   \\  
\hline 				  
    1  &  3s3p$^4$($^3$P)3d			 &  $^5$D$  _4$   &  4.0012  &  4.00972  & 3.9939 &  3.9938  & 4.2027 \\
    2  &  3s3p$^4$($^3$P)3d			 &  $^5$D$  _3$   &  4.0043  &  4.01270  & 3.9969 &  3.9969  & 4.2085 \\
    3  &  3s3p$^4$($^3$P)3d			 &  $^5$D$  _2$   &  4.0081  &  4.01648  & 4.0007 &  4.0007  & 4.2129 \\
    4  &  3s3p$^4$($^3$P)3d			 &  $^5$D$  _1$   &  4.0114  &  4.01983  & 4.0040 &  4.0040  & 4.2157 \\
    5  &  3s3p$^4$($^3$P)3d			 &  $^5$D$  _0$   &  4.0133  &  4.02177  & 4.0059 &  4.0059  & 4.2171 \\
    6  &  3s3p$^4$($^3$P)3d			 &  $^5$F$  _5$   &  4.3242  &  4.32861  & 4.3163 &  4.3109  & 4.4171 \\
    7  &  3s3p$^4$($^3$P)3d			 &  $^5$F$  _4$   &  4.3378  &  4.34200  & 4.3298 &  4.3244  & 4.4317 \\
    8  &  3s3p$^4$($^3$P)3d			 &  $^5$F$  _3$   &  4.3479  &  4.35206  & 4.3398 &  4.3344  & 4.4434 \\
    9  &  3s3p$^4$($^3$P)3d			 &  $^5$F$  _2$   &  4.3550  &  4.35921  & 4.3469 &  4.3415  & 4.4520 \\
   10  &  3s3p$^4$($^3$P)3d			 &  $^5$F$  _1$   &  4.3595  &  4.36380  & 4.3514 &  4.3460  & 4.4578 \\
   11  &  3s3p$^4$($^3$P)3d			 &  $^5$P$  _1$   &  4.3855  &  4.38947  & 4.3766 &  4.3743  & 4.5582 \\
   12  &  3s3p$^4$($^3$P)3d			 &  $^5$P$  _2$   &  4.3958  &  4.39975  & 4.3867 &  4.3843  & 4.5672 \\
   13  &  3s3p$^4$($^3$P)3d			 &  $^5$P$  _3$   &  4.4130  &  4.41691  & 4.4041 &  4.4018  & 4.5805 \\
   14  &  3s3p$^4$($^1$D)3d			 &  $^3$P$  _0$   &  4.4594  &  4.46232  & 4.4394 &  4.4361  & 4.7991 \\
   15  &  3s3p$^4$($^1$D)3d			 &  $^3$P$  _1$   &  4.4696  &  4.47247  & 4.4497 &  4.4465  & 4.8146 \\
   16  &  3s3p$^4$($^1$D)3d			 &  $^3$P$  _2$   &  4.4902  &  4.49293  & 4.4703 &  4.4671  & 4.8313 \\
   17  &  3s3p$^4$($^1$D)3d			 &  $^3$D$  _1$   &  4.6038  &  4.60243  & 4.5861 &  4.5824  & 4.8903 \\
   18  &  3s3p$^4$($^1$D)3d			 &  $^3$D$  _2$   &  4.6139  &  4.61241  & 4.5961 &  4.5924  & 4.9042 \\
   19  &  3s3p$^4$($^1$D)3d			 &  $^3$D$  _3$   &  4.6286  &  4.62692  & 4.6108 &  4.6069  & 4.9120 \\
   20  &  3s3p$^4$($^3$P)3d			 &  $^3$F$  _4$   &  4.6417  &  4.64011  & 4.6238 &  4.6167  & 4.8474 \\
   21  &  3s3p$^4$($^3$P)3d			 &  $^3$F$  _3$   &  4.6664  &  4.66480  & 4.6484 &  4.6414  & 4.8688 \\
   22  &  3s3p$^4$($^3$P)3d			 &  $^3$F$  _2$   &  4.6827  &  4.68102  & 4.6646 &  4.6576  & 4.8846 \\
   23  &  3s3p$^4$($^1$D)3d			 &  $^3$G$  _3$   &  4.7762  &  4.77400  & 4.7549 &  4.7462  & 5.0486 \\
   24  &  3s3p$^4$($^1$D)3d			 &  $^3$G$  _4$   &  4.7799  &  4.77774  & 4.7586 &  4.7498  & 5.0493 \\
   25  &  3s3p$^4$($^1$D)3d			 &  $^3$G$  _5$   &  4.7848  &  4.78256  & 4.7634 &  4.7546  & 5.0501 \\
   26  &  3s3p$^4$($^1$D)3d			 &  $^1$G$  _4$   &  4.9312  &  4.92451  & 4.8936 &  4.8864  & 5.4070 \\
   27  &  3s3p$^4$($^1$D)3d			 &  $^1$D$  _2$   &  4.9796  &  4.97583  & 4.9508 &  4.9458  & 5.4540 \\
   28  &  3s3p$^4$($^1$D)3d			 &  $^3$F$  _3$   &  5.0034  &  4.99912  & 4.9746 &  4.9669  & 5.3014 \\
   29  &  3s3p$^4$($^1$D)3d			 &  $^3$F$  _2$   &  5.0069  &  5.00302  & 4.9781 &  4.9718  & 5.2996 \\
   30  &  3s3p$^4$($^1$D)3d			 &  $^3$F$  _4$   &  5.0113  &  5.00699  & 4.9824 &  4.9746  & 5.3034 \\
   31  &  3s3p$^4$($^3$P)3d			 &  $^3$D$  _3$   &  5.0595  &  5.05232  & 5.0271 &  5.0198  & 5.4789 \\
   32  &  3s3p$^4$($^3$P)3d			 &  $^3$D$  _1$   &  5.0702  &  5.06266  & 5.0380 &  5.0311  & 5.4842 \\
   33  &  3s3p$^4$($^3$P)3d			 &  $^3$D$  _2$   &  5.0736  &  5.06643  & 5.0415 &  5.0349  & 5.4870 \\
   34  &  3s3p$^4$($^1$D)3d			 &  $^1$F$  _3$   &  5.0957  &  5.08866  & 5.0627 &  5.0549  & 5.5296 \\
   35  &  3s3p$^4$($^3$P)3d			 &  $^3$P$  _0$   &  5.1868  &  5.18263  & 5.1511 &  5.1481  & 6.1936 \\
   36  &  3s3p$^4$($^1$D)3d			 &  $^1$P$  _1$   &  5.2046  &  5.18372  & 5.1568 &  5.1505  & 5.3971 \\
   37  &  3s3p$^4$($^3$P)3d			 &  $^3$P$  _2$   &  5.2130  &  5.19858  & 5.1735 &  5.1654  & 6.1728 \\
   38  &  3s3p$^4$($^1$D)3d			 &  $^3$S$  _1$   &  5.2473  &  5.24088  & 5.2249 &  5.2220  & 5.3853 \\
   39  &  3s3p$^4$($^3$P)3d			 &  $^3$F$  _4$   &  5.3425  &  5.33535  & 5.3084 &  5.2926  & 5.8302 \\
   40  &  3s3p$^4$($^3$P)3d			 &  $^3$F$  _3$   &  5.3478  &  5.34036  & 5.3143 &  5.2988  & 5.8459 \\
   41  &  3s3p$^4$($^3$P)3d			 &  $^3$F$  _2$   &  5.3555  &  5.34781  & 5.3222 &  5.3069  & 5.8579 \\
   42  &  3s3p$^4$($^1$S)3d			 &  $^3$D$  _1$   &  5.3837  &  5.37402  & 5.3548 &  5.3441  & 5.4982 \\
   43  &  3s3p$^4$($^1$S)3d			 &  $^3$D$  _2$   &  5.3955  &  5.38619  & 5.3649 &  5.3536  & 5.5119 \\
   44  &  3s3p$^4$($^1$S)3d			 &  $^3$D$  _3$   &  5.4046  &  5.39510  & 5.3753 &  5.3640  & 5.5181 \\
\hline            				    							  
\end{tabular} 

\clearpage
\newpage
\begin{flushleft}
Table 3. Levels of the 3s3p$^4$3d configuration of Ti VII and their excitation energies (in Ryd). 
\end{flushleft}
\begin{tabular}{rllrlrllrl} \hline
Index  & \multicolumn{2}{c}{Configuration/Level} &                   GRASP2  & FAC1      & FAC2   & FAC3     & CIV3   \\  
\hline 
   45  &  3s3p$^4$($^3$P)3d			 &  $^3$P$  _2$   &  5.4373  &  5.42739  & 5.3925 &  5.3835  & 5.8081 \\
   46  &  3s3p$^4$($^3$P)3d			 &  $^3$P$  _1$   &  5.4444  &  5.43443  & 5.3988 &  5.3901  & 5.7975 \\
   47  &  3s3p$^4$($^3$P)3d			 &  $^3$P$  _0$   &  5.4470  &  5.43695  & 5.4010 &  5.3924  & 5.7848 \\
   48  &  3s3p$^4$($^3$P)3d			 &  $^3$D$  _3$   &  5.5042  &  5.49414  & 5.4668 &  5.4596  & 6.0697 \\
   49  &  3s3p$^4$($^1$S)3d			 &  $^1$D$  _2$   &  5.5131  &  5.49906  & 5.4765 &  5.4702  & 5.9065 \\
   50  &  3s3p$^4$($^3$P)3d			 &  $^3$D$  _2$   &  5.5305  &  5.51926  & 5.4927 &  5.4856  & 6.1003 \\
   51  &  3s3p$^4$($^3$P)3d			 &  $^3$D$  _1$   &  5.5420  &  5.53141  & 5.5044 &  5.4971  & 6.1131 \\
   52  &  3s3p$^4$($^3$P)3d			 &  $^1$F$  _3$   &  5.6882  &  5.67577  & 5.6547 &  5.6357  & 5.9107 \\
   53  &  3s3p$^4$($^1$D)3d			 &  $^1$S$  _0$   &  5.7624  &  5.77799  & 5.7464 &  5.7376  & 6.6930 \\
   54  &  3s3p$^4$($^3$P)3d			 &  $^3$P$  _1$   &  5.9878  &  5.96272  & 5.9680 &  5.9435  & 6.1912 \\
   55  &  3s3p$^4$($^3$P)3d			 &  $^1$P$  _1$   &  6.0974  &  5.96816  & 6.0447 &  5.9860  & 6.6301 \\
   56  &  3s3p$^4$($^3$P)3d			 &  $^1$D$  _2$   &  6.1620  &  5.96984  & 6.0985 &  6.0837  & 6.1936 \\
\hline            				    							  
\end{tabular}   					    									
			      				    									
\begin{flushleft}					    									
{\small 								      											       
GRASP2: Present results {\em with} QED effects \\
FAC1: Present results with  4186 levels \\
FAC2: Present results with 38,694 levels \\
FAC3: Present results with 159,162 levels \\
CIV3: Results of Singh {\em et al}   \cite{mm4}  \\ 
}															       
\end{flushleft} 

\clearpage
\newpage
{\small
\begin{flushleft}
{\bf Table 6.} Comparison of oscillator strengths (f- values) for transitions from 3p$^4$ to 3s3p$^5$ and 3p$^3$3d configurations of Ti VII. ($a{\pm}b \equiv a{\times}$10$^{{\pm}b}$). 
\end{flushleft}
\begin{tabular}{rllrrrrrrrrrrrr} \hline
   I$^c$    &    J$^c$&  GRASP   & FAC   &   MCHF    &   CIV3     &   HFR   &    I   &   J   &   GRASP  & FAC   &  MCHF      &	CIV3     &  HFR	    \\
\hline			    	    			
    1  &    6  &   4.1-2  & 4.2-2 &    4.4-2  &    2.6-2   &  4.0-2  &     2  &   44  &   3.9-1  & 3.8-1 &    3.8-1   &   4.3-1  &   3.3-1  \\
    1  &    7  &   1.4-2  & 1.4-2 &    1.5-2  &    1.9-2   &  1.4-2  &     2  &   45  &   1.7-2  & 1.7-2 &    2.1-2   &   .....  &   1.2-2  \\
    1  &    9  &   6.5-4  & 6.7-4 &    6.4-4  &    .....   &  6.3-4  &     3  &    7  &   5.5-2  & 5.6-2 &    5.8-2   &   3.1-2  &   5.4-2  \\
    1  &   11  &   9.5-5  & 1.0-4 &    8.9-5  &    .....   &  3.8-5  &     3  &    9  &   2.1-5  & ..... &    .....   &   .....  &   7.9-6  \\
    1  &   12  &   2.1-4  & 2.2-4 &    1.9-4  &    .....   &  7.6-5  &     3  &   11  &   1.1-4  & 1.1-4 &    1.0-4   &   .....  &   6.2-5  \\
    1  &   13  &   1.1-4  & 1.2-4 &    1.0-4  &    .....   &  2.6-5  &     3  &   17  &   3.2-5  & 2.1-5 &    4.8-5   &   2.0-2  &   1.1-3  \\
    1  &   15  &   2.0-4  & 1.9-4 &    1.9-4  &    4.6-3   &  3.5-4  &     3  &   31  &   1.3-2  & 1.5-2 &    1.0-3   &   7.5-6  &   4.4-3  \\
    1  &   16  &   2.2-6  & 8.5-7 &    8.3-6  &    2.8-2   &  7.4-4  &     3  &   32  &   1.1-2  & 1.0-2 &    1.9-2   &   2.3-2  &   6.2-2  \\
    1  &   17  &   3.7-5  & 3.6-5 &    3.3-5  &    2.0-4   &  3.6-5  &     3  &   37  &   5.8-2  & 3.6-2 &    1.1-1   &   3.1-1  &   7.6-2  \\
    1  &   18  &   2.3-5  & 2.4-5 &    2.2-5  &    .....   &  8.7-6  &     3  &   39  &   7.6-1  & 7.3-1 &    6.0-1   &   1.0-0  &   4.9-1  \\
    1  &   20  &   5.6-5  & 5.8-5 &    5.6-5  &    .....   &  3.6-5  &     3  &   41  &   2.9-1  & 3.4-1 &    3.9-1   &   .....  &   4.6-1  \\
    1  &   22  &   3.9-5  & 3.9-5 &    3.6-5  &    .....   &  9.1-6  &     3  &   44  &   1.6-0  & 1.6-0 &    1.6-0   &   1.7-0  &   1.4-0  \\
    1  &   26  &   2.0-4  & 2.0-4 &    1.7-4  &    .....   &  1.8-4  &     4  &    6  &   4.8-4  & 5.0-4 &    .....   &   .....  &   6.0-4  \\
    1  &   29  &   4.9-4  & 5.0-4 &    4.5-4  &    .....   &  9.1-4  &     4  &    7  &   1.4-6  & ..... &    .....   &   .....  &   1.6-8  \\
    1  &   30  &   2.5-6  & 2.9-6 &    5.3-6  &    .....   &  8.7-5  &     4  &    9  &   7.0-2  & 7.1-2 &    7.1-2   &   3.4-2  &   5.5-2  \\
    1  &   31  &   7.8-3  & 7.6-3 &    5.5-3  &    3.7-6   &  5.4-6  &     4  &   11  &   3.2-7  & 2.6-7 &    .....   &   .....  &   2.3-7  \\
    1  &   32  &   1.2-4  & 2.3-4 &    9.5-4  &    2.4-4   &  2.4-4  &     4  &   12  &   7.2-7  & 9.9-7 &    .....   &   .....  &   3.6-7  \\
    1  &   33  &   1.6-2  & 1.2-2 &    1.8-2  &    1.4-4   &  7.4-4  &     4  &   13  &   3.1-7  & 4.1-7 &    .....   &   .....  &   8.0-8  \\
    1  &   34  &   7.4-3  & 1.1-2 &    3.1-4  &    5.6-3   &  5.8-3  &     4  &   15  &   2.4-5  & 2.3-5 &    1.8-5   &   .....  &   3.3-5  \\
    1  &   35  &   5.1-3  & 5.1-3 &    4.8-3  &    1.9-2   &  4.9-2  &     4  &   16  &   5.3-6  & 5.3-6 &    4.6-6   &   .....  &   2.5-6  \\
    1  &   36  &   2.0-3  & 2.0-3 &    1.6-3  &    .....   &  8.7-4  &     4  &   17  &   5.5-6  & 5.8-6 &    3.8-6   &   .....  &   4.1-6  \\
    1  &   37  &   3.7-1  & 3.8-1 &    3.0-1  &    3.1-1   &  3.1-1  &     4  &   18  &   1.2-5  & 1.3-5 &    1.1-5   &   .....  &   1.7-5  \\
    1  &   38  &   6.9-1  & 6.9-1 &    7.2-1  &    7.7-1   &  6.9-1  &     4  &   20  &   9.1-5  & 9.3-5 &    7.1-5   &   .....  &   2.8-5  \\
    1  &   39  &   9.0-2  & 6.9-2 &    1.2-1  &    2.6-1   &  8.9-2  &     4  &   22  &   3.2-4  & 3.2-4 &    2.7-4   &   .....  &   8.9-5  \\
    1  &   41  &   2.0-2  & 2.2-2 &    4.1-2  &    .....   &  4.9-2  &     4  &   26  &   1.3-2  & 1.2-2 &    1.1-2   &   7.4-2  &   2.6-2  \\
    1  &   42  &   1.3-0  & 1.3-0 &    1.3-0  &    1.4-0   &  1.2-0  &     4  &   29  &   1.5-4  & 1.6-4 &    1.4-4   &   .....  &   4.3-4  \\
    1  &   43  &   2.1-1  & 2.1-1 &    2.1-1  &    2.6-1   &  1.8-1  &     4  &   30  &   3.9-4  & 3.8-4 &    3.6-4   &   .....  &   2.8-4  \\
    1  &   44  &   1.3-2  & 1.3-2 &    1.2-2  &    1.7-2   &  1.1-2  &     4  &   31  &   6.0-4  & 6.7-4 &    2.2-4   &   .....  &   5.9-4  \\
    1  &   45  &   2.3-4  & 2.4-4 &    .....  &    .....   &  5.8-4  &     4  &   32  &   2.4-4  & 2.1-4 &    6.8-4   &   .....  &   2.3-4  \\
    1  &   47  &   1.1-3  & 1.1-3 &    .....  &    .....   &  2.4-3  &     4  &   33  &   5.6-6  & ..... &    .....   &   .....  &   8.2-7  \\
    2  &    6  &   2.3-2  & 2.3-2 &    2.4-2  &    1.4-2   &  2.3-2  &     4  &   34  &   4.6-4  & 4.7-4 &    1.9-4   &   .....  &   7.6-5  \\
    2  &    7  &   1.4-2  & 1.4-2 &    1.5-2  &    1.1-2   &  1.4-2  &     4  &   35  &   2.3-4  & 2.4-4 &    2.5-4   &   .....  &   1.3-3  \\
    2  &    8  &   1.9-2  & 1.9-2 &    2.0-2  &    1.1-2   &  1.8-2  &     4  &   36  &   4.9-4  & 4.9-4 &    1.3-4   &   7.1-2  &   1.6-2  \\
    2  &    9  &   4.8-5  & 4.9-5 &    .....  &    .....   &  2.8-5  &     4  &   37  &   5.8-4  & 1.2-4 &    .....   &   .....  &   3.7-4  \\
    2  &   10  &   1.3-4  & 1.3-4 &    1.2-4  &    .....   &  5.3-5  &     4  &   38  &   4.2-3  & 4.3-3 &    .....   &   .....  &   5.0-3  \\
    2  &   11  &   1.3-4  & 1.3-4 &    1.2-4  &    .....   &  1.4-4  &     4  &   39  &   1.2-1  & 1.4-1 &    1.7-1   &   .....  &   2.2-1  \\
    2  &   12  &   9.9-7  & 1.2-6 &    .....  &    .....   &  9.0-7  &     4  &   41  &   3.8-1  & 3.6-1 &    3.4-1   &   5.9-1  &   2.8-1  \\
    2  &   15  &   3.1-7  & 1.0-6 &    .....  &    2.3-2   &  6.2-4  &     4  &   42  &   5.2-4  & 5.2-4 &    .....   &   .....  &   1.1-3  \\
    2  &   17  &   1.2-4  & 1.2-4 &    1.2-4  &    5.0-3   &  4.2-4  &     4  &   43  &   4.6-3  & 4.5-3 &    .....   &   .....  &   2.8-3  \\
    2  &   18  &   1.1-5  & 1.0-5 &    1.3-5  &    .....   &  1.4-6  &     4  &   44  &   8.7-4  & 8.2-4 &    .....   &   .....  &   1.2-3  \\
    2  &   19  &   8.2-5  & 8.3-5 &    .....  &    .....   &  2.2-5  &     4  &   45  &   7.9-1  & 7.7-1 &    7.7-1   &   8.9-1  &   6.9-1  \\
    2  &   26  &   1.4-4  & 1.4-4 &    6.7-5  &    .....   &  8.2-5  &     4  &   47  &   1.4-0  & 1.4-0 &    .....   &   1.5-0  &   1.3-0  \\
    2  &   28  &   3.2-3  & 3.1-3 &    2.1-3  &    4.5-4   &  1.2-3  &     5  &    7  &   3.4-4  & 3.5-4 &    .....   &   .....  &   3.7-4  \\
    2  &   30  &   1.2-4  & 1.1-4 &    8.5-5  &    .....   &  4.9-4  &     5  &    9  &   7.9-3  & 8.6-3 &    9.4-3   &   4.3-4  &   8.5-3  \\
    2  &   31  &   8.7-4  & 7.1-4 &    1.5-3  &    2.4-6   &  2.2-3  &     5  &   11  &   1.3-7  & ..... &    .....   &   .....  &   9.6-8  \\
    2  &   32  &   2.0-3  & 2.1-3 &    7.0-4  &    5.9-3   &  1.2-2  &     5  &   17  &   2.4-5  & 2.7-5 &    2.1-5   &   .....  &   3.0-7  \\
    2  &   33  &   2.2-2  & 2.3-2 &    6.3-3  &    8.4-5   &  2.5-4  &     5  &   31  &   4.9-5  & 2.5-5 &    .....   &   .....  &   1.7-5  \\
    2  &   34  &   5.9-4  & 4.9-6 &    1.2-2  &    2.7-2   &  4.7-2  &     5  &   32  &   7.2-4  & 7.6-4 &    .....   &   .....  &   2.7-6  \\
    2  &   37  &   1.2-1  & 9.8-2 &    1.6-1  &    3.1-1   &  1.3-1  &     5  &   37  &   6.8-4  & 1.0-3 &    .....   &   .....  &   1.3-2  \\
    2  &   38  &   3.2-1  & 3.1-1 &    3.1-1  &    4.2-1   &  3.1-1  &     5  &   39  &   2.9-2  & 3.5-2 &    3.8-2   &   .....  &   4.1-2  \\
    2  &   39  &   2.8-1  & 2.8-1 &    2.1-1  &    2.6-1   &  1.8-1  &     5  &   41  &   1.3-1  & 1.3-1 &    1.0-1   &   1.3-1  &   9.1-2  \\
    2  &   40  &   3.1-1  & 3.0-1 &    3.1-1  &    3.5-1   &  3.0-1  &     5  &   44  &   1.7-5  & ..... &    .....   &   .....  &   2.5-3  \\
    2  &   41  &   8.7-2  & 1.0-1 &    1.1-1  &    .....   &  1.4-1  &     5  &   48  &   2.7-0  & 2.6-0 &    2.7-0   &   3.3-0  &   2.5-0  \\
    2  &   43  &   1.2-0  & 1.2-0 &    1.2-0  &    1.3-0   &  1.1-0  &        &       & 	 &  &		 &	    &	       \\
 \hline				  			     	 	 					       
\end{tabular}
}						   		   					       
\begin {flushleft}												        	   
\begin{tabbing} 												       
aaaaaaaaaaaaaaaaaaaaaaaaaaaaaaaaaaaa\= \kill									       
$c$: Level indices for transitions correspond to those of Table 4, \\available online at stacks.iop.org/PhysScr/vol/number/mmedia \\
GRASP: Present results with the {\sc grasp} code   \\								       
FAC: Present results with the {\sc fac} code \\ 								       
MCHF: Results of Froese-Fischer {\em et al} \cite{cff1} with the MCHF code  \\ 
CIV3: Results of Singh {\em et al}  \cite{mm4} with the CIV3 code  \\ 
HFR:  Results of Biemont  \cite{eb} with the HFR code \\ 
\end{tabbing}													       
\end {flushleft}												       
\end{document}